\documentclass[
reprint,
amsmath,
amssymb,
aps,
prc,
superscriptaddress
]{revtex4-2}

\usepackage{graphicx}
\usepackage{dcolumn}
\usepackage{bm}
\usepackage{comment}
\usepackage{CJK} 
\usepackage{amsmath}
\begin{document}

%\title{Lifetime measurements of low-lying octupole states in $^{224}$Ra}
\title{Independent Validation of Octupole Collectivity in $^{224}$Ra Through Lifetime Measurements of Low-lying Negative-parity States}

\author{D. White}
\email{dylan.white@uws.ac.uk}
\affiliation{Faculty of Computing, Sciences and Engineering,
University of the West of Scotland, Paisley, PA1 2BE, United Kingdom}

\author{D. O'Donnell}
\email{david.odonnell@uws.ac.uk}
\affiliation{Faculty of Computing, Sciences and Engineering,
University of the West of Scotland, Paisley, PA1 2BE, United Kingdom}

% Resources for authorship from combined S2189 and S2319 beamtime
% Beamtime was June 17-23 2024
% https://grsi.wiki.triumf.ca/w/index.php?title=S2189,_S2319
% https://triumfoffice365-my.sharepoint.com/:x:/g/personal/aavaa_triumf_ca/EfEklLuPbXtMocxC4kiUvGkBuF_C77G6C1dPA8TStJl9RQ?e=8BfWUG

% Leading authors who played major role in spectrometer setup and operation
\author{A. Avaa}
\affiliation{TRIUMF, 4004 Wesbrook Mall, Vancouver, British Columbia V6T 2A3, Canada}
\author{J.~R. Murias}
\affiliation{TRIUMF, 4004 Wesbrook Mall, Vancouver, British Columbia V6T 2A3, Canada}
%Present: DIAMOND
\author{S. Murillo Morales}
\affiliation{TRIUMF, 4004 Wesbrook Mall, Vancouver, British Columbia V6T 2A3, Canada}
\author{R.~Umashankar}
\affiliation{TRIUMF, 4004 Wesbrook Mall, Vancouver, British Columbia V6T 2A3, Canada}
\affiliation{Department of Physics and Astronomy, University of British Columbia, Vancouver, BC V6T 1Z4, Canada}
\author{V.~Vedia}
\affiliation{TRIUMF, 4004 Wesbrook Mall, Vancouver, British Columbia V6T 2A3, Canada}
%Present: CERN

% Alphabetical
\author{C. Andreoiu}
\affiliation{Department of Chemistry, Simon Fraser University, 8888 University Drive, Burnaby, British Columbia V5A 1S6, Canada}
\author{D.~W. Annen}
\affiliation{Department of Physics, Simon Fraser University, 8888 University Drive, Burnaby, British Columbia V5A 1S6, Canada}
\author{A.~D. Ayangeakaa}
\affiliation{University of North Carolina, Chapel Hill, North Carolina 27599, USA}
\affiliation{Triangle Universities Nuclear Laboratory, Duke University, Durham, North Carolina 27708, USA}
\author{G.~C. Ball}
\affiliation{TRIUMF, 4004 Wesbrook Mall, Vancouver, British Columbia V6T 2A3, Canada}
\author{V. Bildstein}
\affiliation{Department of Physics, University of Guelph, Guelph, ON, N1G 2W1, Canada}
\author{M. Bowry}
\affiliation{Faculty of Computing, Sciences and Engineering, University of the West of Scotland, Paisley, PA1 2BE, United Kingdom}
\author{I. Dillmann}
\affiliation{TRIUMF, 4004 Wesbrook Mall, Vancouver, British Columbia V6T 2A3, Canada}
\affiliation{Department of Physics and Astronomy, University of Victoria, Victoria, BC, V8P 5C2, Canada}
\author{E.~G. Fuakye}
\affiliation{Department of Physics, University of Regina, Regina, SK, S4S 0A2, Canada}
\author{L.~P. Gaffney}
\affiliation{Oliver Lodge Laboratory, University of Liverpool, Liverpool, L69 7ZE, United Kingdom}
\author{A.~B. Garnsworthy}
\affiliation{TRIUMF, 4004 Wesbrook Mall, Vancouver, British Columbia V6T 2A3, Canada}
\affiliation{Department of Physics and Astronomy, University of Victoria, Victoria, BC, V8P 5C2, Canada}
\author{P.~E. Garrett}
\affiliation{Department of Physics, University of Guelph, Guelph, ON, N1G 2W1, Canada}
\author{E.~D. Geerlof}
\affiliation{TRIUMF, 4004 Wesbrook Mall, Vancouver, British Columbia V6T 2A3, Canada}
\author{S. Georges}
\affiliation{TRIUMF, 4004 Wesbrook Mall, Vancouver, British Columbia V6T 2A3, Canada}
\author{A.~L. Grimes}
\affiliation{TRIUMF, 4004 Wesbrook Mall, Vancouver, British Columbia V6T 2A3, Canada}
\author{G.~F. Grinyer}
\affiliation{Department of Physics, University of Regina, Regina, SK, S4S 0A2, Canada}
\author{G.~Hackman}
\affiliation{TRIUMF, 4004 Wesbrook Mall, Vancouver, British Columbia V6T 2A3, Canada}
\affiliation{Department of Chemistry, Simon Fraser University, 8888 University Drive, Burnaby, British Columbia V5A 1S6, Canada}
\author{P.~M. Jones}
\affiliation{iThemba LABS, National Research Foundation, Faure, 7131, South Africa}
\author{J. Liu}
\affiliation{Department of Physics, University of Regina, Regina, SK, S4S 0A2, Canada}
%Present: TRIUMF, 4004 Wesbrook Mall, Vancouver, British Columbia V6T 2A3, Canada
%Present: AND Department of Physics and Astronomy, University of Victoria, Victoria, BC, V8P 5C2, Canada
\author{B. Olaizola}
\affiliation{Instituto de Estructura de la Materia, CSIC, E-28006 Madrid, Spain}
\author{S.~D. Olorunfunmi}
\affiliation{Institute of Physics, University of S\~ao Paulo, Brazil}
\author{F. Rowntree}
\affiliation{Oliver Lodge Laboratory, University of Liverpool, Liverpool, L69 7ZE, United Kingdom}
\author{N.~K. Syeda}
\affiliation{Department of Physics, Simon Fraser University, 8888 University Drive, Burnaby, British Columbia V5A 1S6, Canada}
\author{D. Shah}
\affiliation{Department of Physics, University of Regina, Regina, SK, S4S 0A2, Canada}
\author{P. Spagnoletti}
\affiliation{Department of Chemistry, Simon Fraser University, 8888 University Drive, Burnaby, British Columbia V5A 1S6, Canada}
\affiliation{Oliver Lodge Laboratory, University of Liverpool, Liverpool, L69 7ZE, United Kingdom}
\author{C.~E. Svensson}
\affiliation{Department of Physics, University of Guelph, Guelph, ON, N1G 2W1, Canada}
\affiliation{TRIUMF, 4004 Wesbrook Mall, Vancouver, British Columbia V6T 2A3, Canada}

\begin{CJK*}{UTF8}{gbsn} 
\author{F.~Wu (吴桐安)}
\affiliation{Department of Chemistry, Simon Fraser University, 8888 University Drive, Burnaby, British Columbia V5A 1S6, Canada}

\date{\today}

\begin{abstract}
The nucleus $^{224}$Ra is a key benchmark for octupole deformation and for theoretical descriptions of enhanced Schiff moments in reflection-asymmetric nuclei. While Coulomb-excitation measurements have established strong octupole collectivity in $^{224}$Ra, theoretical models predict that its intrinsic electric-dipole moment should be strongly quenched by a cancellation between macroscopic and microscopic contributions. Direct fast-timing measurements of the low-lying $J^\pi = 1^-_1$ and $3^-_1$ states populated following the $\beta$-decay of $^{224}$Fr at TRIUMF-ISAC were performed. Using the LaBr$_3$(Ce) detectors of the GRIFFIN array, mean lifetimes of $\tau(1^-_1) = 444(6)$~ps and $\tau(3^-_1) = 460(18)$~ps were obtained. The corresponding reduced transition probabilities agree with values inferred from Coulomb excitation, but are determined with substantially improved precision. These results provide an independent validation of the electromagnetic matrix elements associated with octupole collectivity in $^{224}$Ra and confirm a strongly-quenched intrinsic dipole moment of $D_0 \simeq 0.032~e\mathrm{fm}$. The present measurements therefore provide a stringent experimental benchmark for nuclear-structure models used in the interpretation of Schiff moments and future searches for non-zero electric dipole moments.
\end{abstract}

\maketitle
\end{CJK*}

%\section{Introduction}

Fundamentally, the scientific method relies upon the independent verification of experimental results. In many areas of contemporary science, landmark discoveries are routinely subjected to repeated measurements using different techniques and instrumentation before they become firmly established. Such studies are comparatively rare in experimental nuclear physics, where measurements often require access to unique radioactive-beam facilities and highly oversubscribed experimental programmes. Consequently, influential results may remain largely unchallenged for many years despite their importance for the development of the field.

One of the most significant developments in nuclear structure physics in recent years has been the emergence of compelling evidence for stable octupole deformation in several neutron-rich actinide nuclei~\cite{Gaffney,Butler_2016,Robledo2013,Agbemava,Butler2020,Chishti2020,Nomura_2021,Nomura2025}. Reflection-asymmetric nuclear shapes arise from strong octupole correlations between orbitals differing by three units of orbital angular momentum and total angular momentum~\cite{Butler_2016}, producing characteristic enhancements of electric-octupole transition strengths and, typically, enhanced intrinsic electric dipole moments. These phenomena are of particular interest because they generate nuclei with enhanced sensitivity to parity- and time-reversal-violating interactions, making them prime candidates for searches for permanent electric dipole moments~\cite{Auerbach1996,Chupp,Flambaum2020}.

Among the nuclei exhibiting the strongest signatures of octupole collectivity, $^{224}$Ra occupies a particularly prominent position. In a Coulomb-excitation experiment, Gaffney {\it et al}.~\cite{Gaffney} reported direct evidence for a large intrinsic octupole moment in $^{224}$Ra, consistent with a stable reflection-asymmetric shape. That work represented a major advance in the field and has subsequently provided much of the experimental motivation for ongoing programmes investigating octupole-deformed nuclei and their relevance to fundamental-symmetry tests.

While enhanced electric-dipole transitions are frequently associated with reflection-asymmetric nuclei, due to the separation of the centres of mass and charge, the situation in $^{224}$Ra is considerably more subtle. Theoretical studies have long predicted~\cite{Butler1991,Butler1996} that the macroscopic and microscopic contributions to the intrinsic electric dipole moment, $D_0$, should nearly cancel, resulting in unusually weak $E1$ transition strengths despite the presence of strong octupole collectivity. Consequently, measurements of the lifetimes of the low-lying negative-parity states provide not only a test of the octupole deformation itself, but also a stringent benchmark of the microscopic mechanisms responsible for the observed suppression of the intrinsic dipole moment.

Despite its importance, relatively few independent measurements have been performed to verify the electromagnetic matrix elements extracted from the Coulomb-excitation analysis. Lifetimes of excited nuclear states provide a complementary and largely model-independent route to determining reduced transition probabilities and therefore offer an important cross-check of the inferred collective behaviour. In particular, measurements of the low-lying negative-parity states in $^{224}$Ra directly probe the octupole degree of freedom responsible for the reflection-asymmetric structure identified in Coulomb-excitation studies.

In the present work, lifetimes of the lowest-lying $J^\pi = 1^-$ and $3^-$ states in $^{224}$Ra have been measured using fast-timing techniques following the $\beta$-decay of $^{224}$Fr. These measurements provide an independent determination of the corresponding electromagnetic transition strengths and constitute one of the few direct experimental tests of the conclusions drawn from the seminal Coulomb-excitation work. The extracted lifetimes are found to be in excellent agreement with the values inferred from previous studies, providing strong confirmation of the octupole collectivity and reflection-asymmetric structure of $^{224}$Ra. More broadly, the present results demonstrate the value of independent measurements in establishing the reliability of benchmark results that underpin contemporary nuclear structure physics. Additionally, reproducing both the large octupole collectivity and the suppressed intrinsic dipole moment provides an important validation of the nuclear structure models used to estimate Schiff moments and parity-violating observables in octupole-deformed nuclei.

%\section{Experimental Details}
This experiment was conducted at TRIUMF, Canada's particle accelerator centre. Here, the $520$ MeV cyclotron was used to provide a proton beam to the Isotope Separator and Accelerator (ISAC) I area. The proton beam was then incident upon a combined target/surface-ion source of uranium carbide, UC$_{\mathrm{x}}$ with a rhenium surface-ion source. The Isotope Separation On-Line (ISOL) technique was used to extract the $^{224}$Fr ions from the rest of the ISOL products.

\begin{figure}
    \centering
    \includegraphics[width=0.65\linewidth]{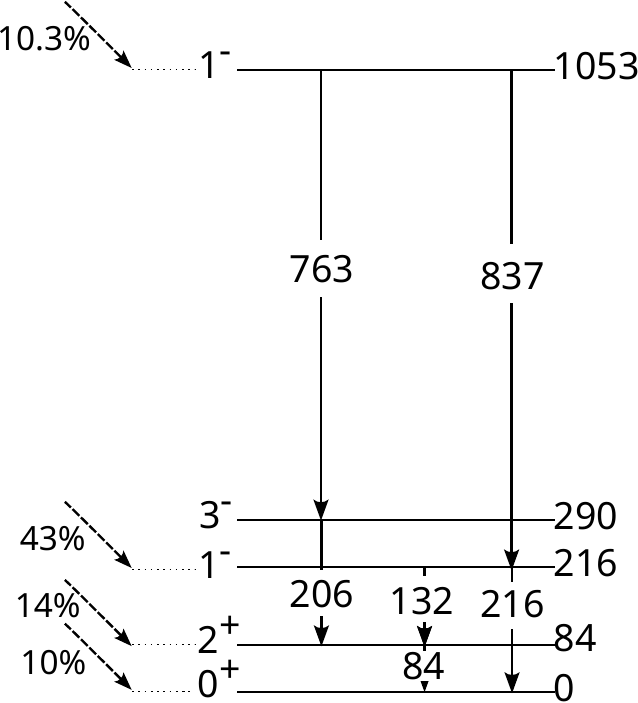}
    \caption{A level scheme showing low-energy states and transitions in $^{224}$Ra which are of relevance to the present work. States are represented by horizontal lines with $J^\pi$ and excitation energy values, in keV, given on the left- and right-hand sides, respectively, of these lines. Transitions between states are indicated by vertical arrows with the energy, in keV, indicated. The diagonal dashed lines on the left-hand side of the figure indicate the feeding of the level via the $\beta$-decay of the $^{224}$Fr parent nuclei~\cite{A224}. }
    \label{levelscheme}
\end{figure}

The $^{224}$Fr ions were implanted into a thin Mylar moving tape collector situated at the centre of the Gamma Ray Infrastructure For Fundamental Investigations of Nuclei (GRIFFIN) array~\cite{GRIFFIN}. There, the implanted $^{224}$Fr nuclei decayed to populate states in $^{224}$Ra. The relevant states for this study, and their feeding intensities, are shown in Figure~\ref{levelscheme}. The tape was moved every 17~minutes, corresponding to approximately five half-lives for $^{224}$Fr, removing activity from the centre of the array to a shielded area at the end of the beam line. This ensured the influence of background radiation from the subsequent decay of the ground state of $^{224}$Ra was minimised.

\begin{figure}
    \centering
    \includegraphics[width=1\linewidth]{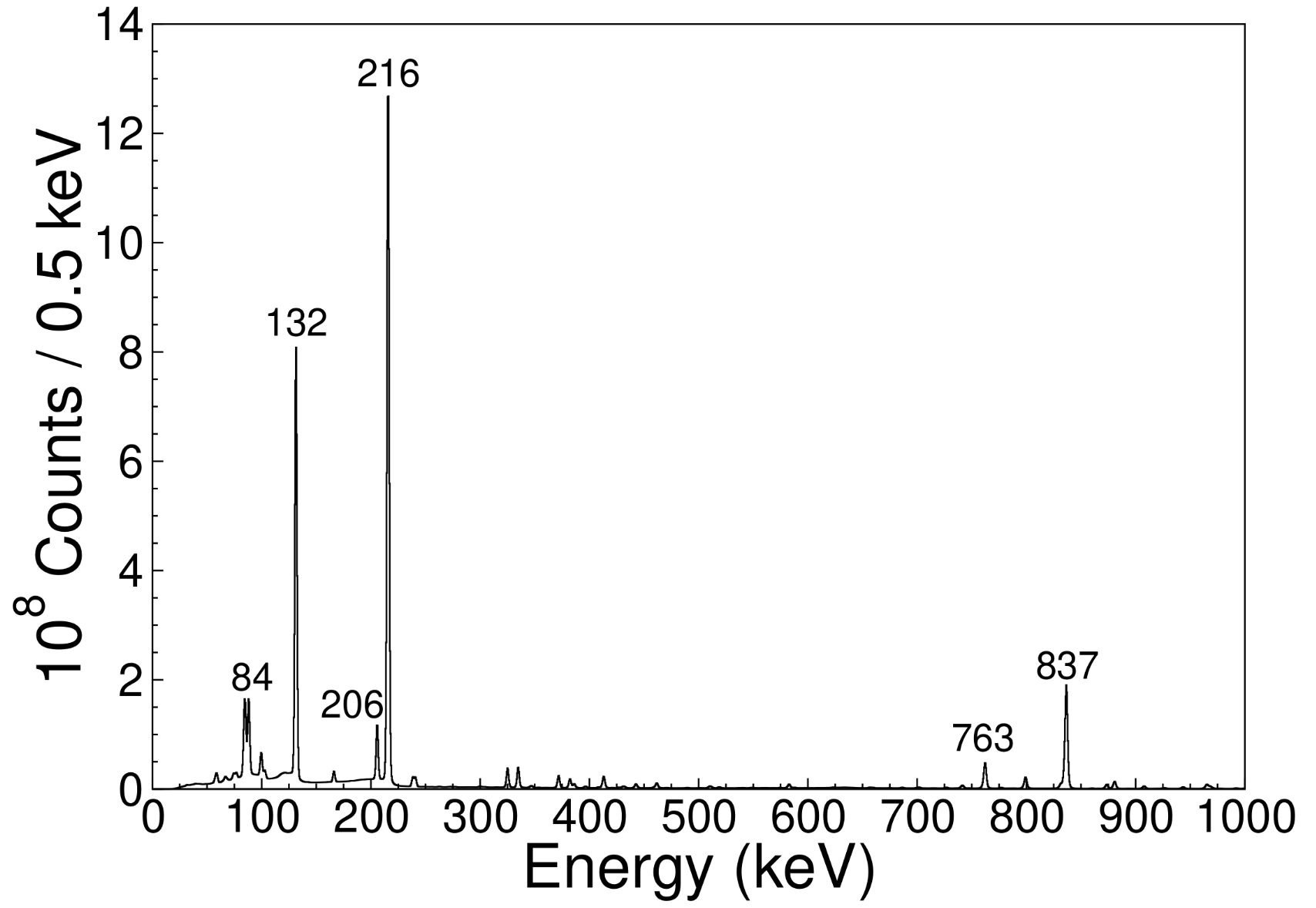}
    \caption{An energy spectrum of $\gamma$ rays detected with the HPGe detectors of the GRIFFIN array showing the most intense transitions between the levels populated by the $\beta$ decay of $^{224}\mathrm{Fr}$ to $^{224}\mathrm{Ra}$. The labels next to the peaks indicate the energies of the transitions. This spectrum was generated using both Compton suppression and add back.}
    \label{griffinSpec}
\end{figure}

\begin{figure}
    \centering
    \includegraphics[width=1\linewidth]{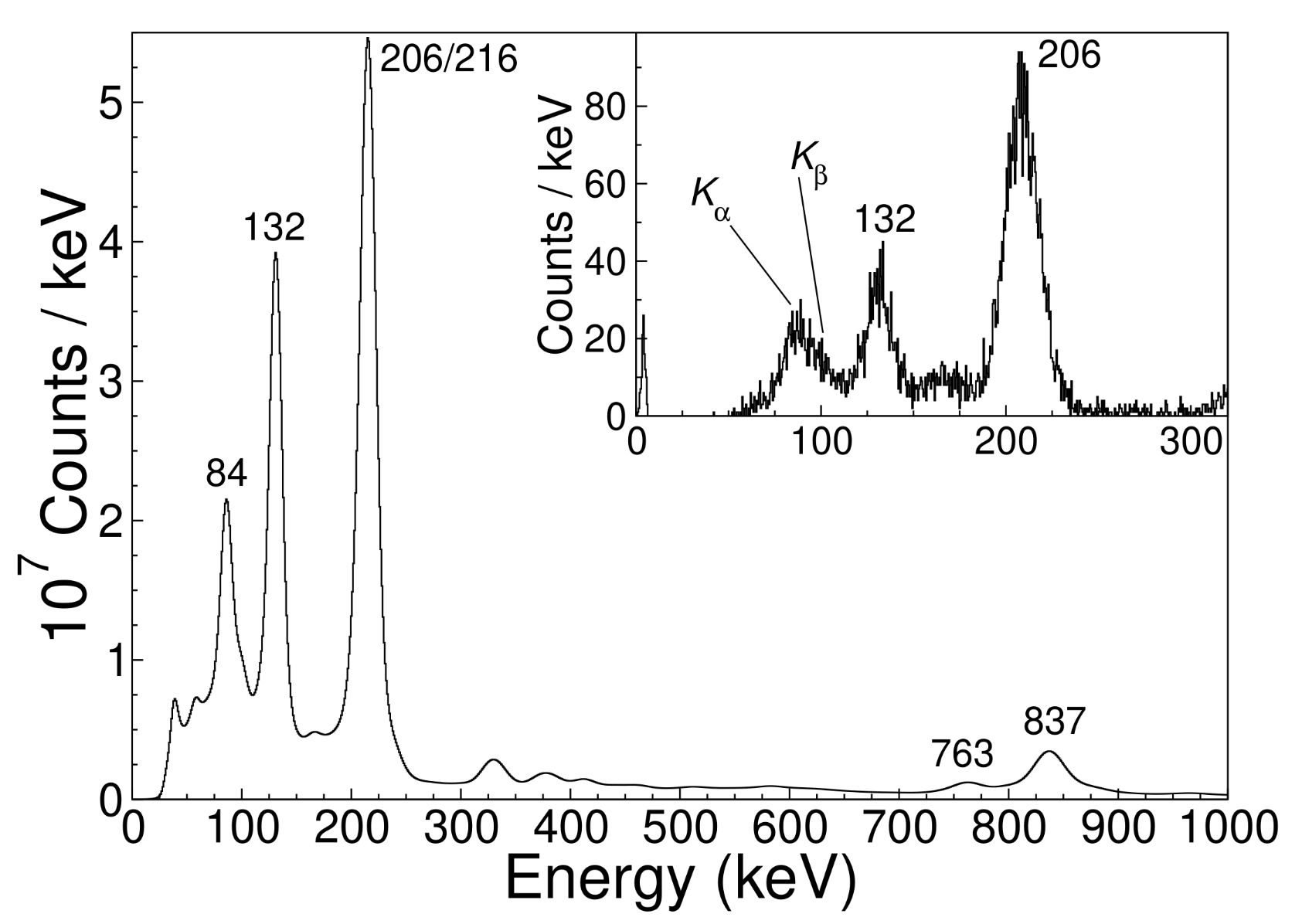}
    \caption{A Compton-suppressed energy spectrum of $\gamma$ rays detected with the LaBr$_3$(Ce) detectors of the GRIFFIN array showing the most intense transitions between the levels populated by the $\beta$ decay of $^{224}\mathrm{Fr}$ to $^{224}\mathrm{Ra}$. The inset shows a spectrum corresponding to events observed in LaBr$_3$(Ce) detectors in coincidence with the detection of 763~keV $\gamma$ rays in other LaBr$_3$(Ce) detectors and with 84~keV $\gamma$ rays detected in HPGe detectors and 67~keV and 80~keV conversion electrons in PACES. }
    \label{labrSpec}
\end{figure}

\begin{figure}
    \centering
    \includegraphics[width=1\linewidth]{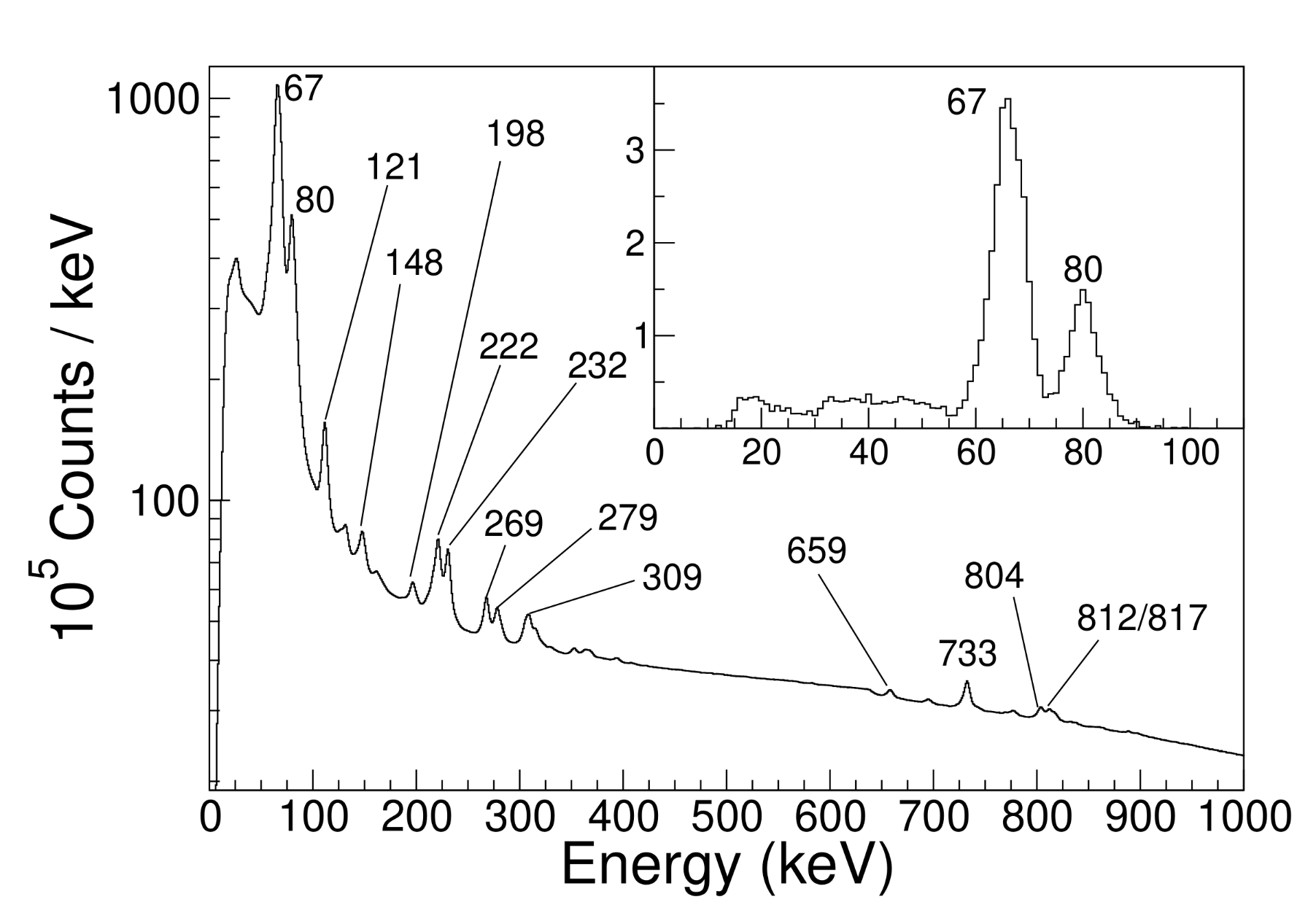}
    \caption{An energy spectrum showing electrons, generated through internal conversion (IC) processes, detected with the PACES detectors. The text labels indicate the energy of the electron peaks. The inset in the top-right shows IC electrons in coincidence with the detection of $206$~keV transitions in the LaBr$_3$(Ce).}
    \label{pacesSpec}
\end{figure}

For this experiment, the GRIFFIN array consisted of fifteen high-purity germanium clover detectors (HPGe), eight cerium-doped lanthanum bromide detectors (LaBr$_3$(Ce)), the Pentagonal Array for Conversion Electron Spectroscopy (PACES) and the Zero-Degree Spectrometer. The Compton suppression system of GRIFFIN surrounded both the HPGe and LaBr$_3$(Ce) detectors. These detectors were calibrated using standard sources of $^{152}$Eu, $^{133}$Ba and $^{60}$Co. Energy spectra of the decay of $^{224}$Fr to $^{224}$Ra from each of HPGe, LaBr$_3$(Ce) and PACES detectors are shown in Figures~\ref{griffinSpec}, \ref{labrSpec} and \ref{pacesSpec}, respectively. 

The LaBr$_3$(Ce) detectors were used to perform the fast-timing measurements using the methodology outlined in Refs.~\cite{Regis2010,Regis2025}. This then allowed for the measurement of mean lifetimes for the states of interest. The signals from these detectors were first sent to linear fan-in/fan out modules. These allowed for the signals to be split, without loss, with one going directly to the GRIF-16 digitizer and the second going to Ortec 935 constant fraction discriminators (CFDs) and Ortec 566 time-to-amplitude convertors (TACs). The signals going directly to the digitizers allowed for the energy of the incident $\gamma$-ray to be measured, while the CFDs and TACs constituted the timing branch.

The HPGe detectors were used primarily to gate on transitions coincident with those feeding and decaying from the state of interest. Utilising the HPGe detectors like this effectively isolated the decay path and helped to reduce the influence of background and uncorrelated events on the lifetime measurement.

The transitions feeding ($E_f$) and decaying ($E_d$) from each state of interest are shown in Figure~\ref{levelscheme}. Signals corresponding to the detection of these events were used to provide the Start and Stop signals of the TACs. The height of the output pulses from each TAC was proportional to the difference in time between the Start and Stop signals. TAC pulse-height histograms were generated for two configurations: the delayed configuration, in which $E_f$ started the TAC and $E_d$ provided the Stop, and the anti-delayed configuration in which the signals were reversed. The TAC histogram of the anti-delayed configuration was then inverted about $t=0$ and summed with the TAC histogram of the delayed configuration, in accordance with the procedure outlined in Refs.~\cite{Karayonchev2017,Ansari,Karayonchev2019}. The resulting combined spectra were subsequently fitted with a function consisting of a Gaussian convoluted with an exponential function of the form  

\begin{align*}
f(x) = {}&\sqrt{\pi}\,\frac{A\lambda}{2}
\exp\left[
\frac{\lambda}{2}
\left(2\mu+\lambda\sigma^2-2x\right)
\right] \\
&\times\operatorname{erfc}\left(
\frac{\mu+\lambda\sigma^2-x}{\sqrt{2}\sigma}
\right)
\end{align*}
\noindent where $A,\mu$ and $\sigma$ correspond to the amplitude, centroid and standard deviation, respectively, of the Gaussian representing the prompt time distribution. Each of the TAC histograms were fitted using this function and the maximum likelihood technique in ROOT~\cite{ROOT}, which then allowed for the measurement of a mean lifetime, $\tau = \frac{1}{\lambda}$.

\begin{figure*}
    \centering
    \includegraphics[width=1\linewidth]{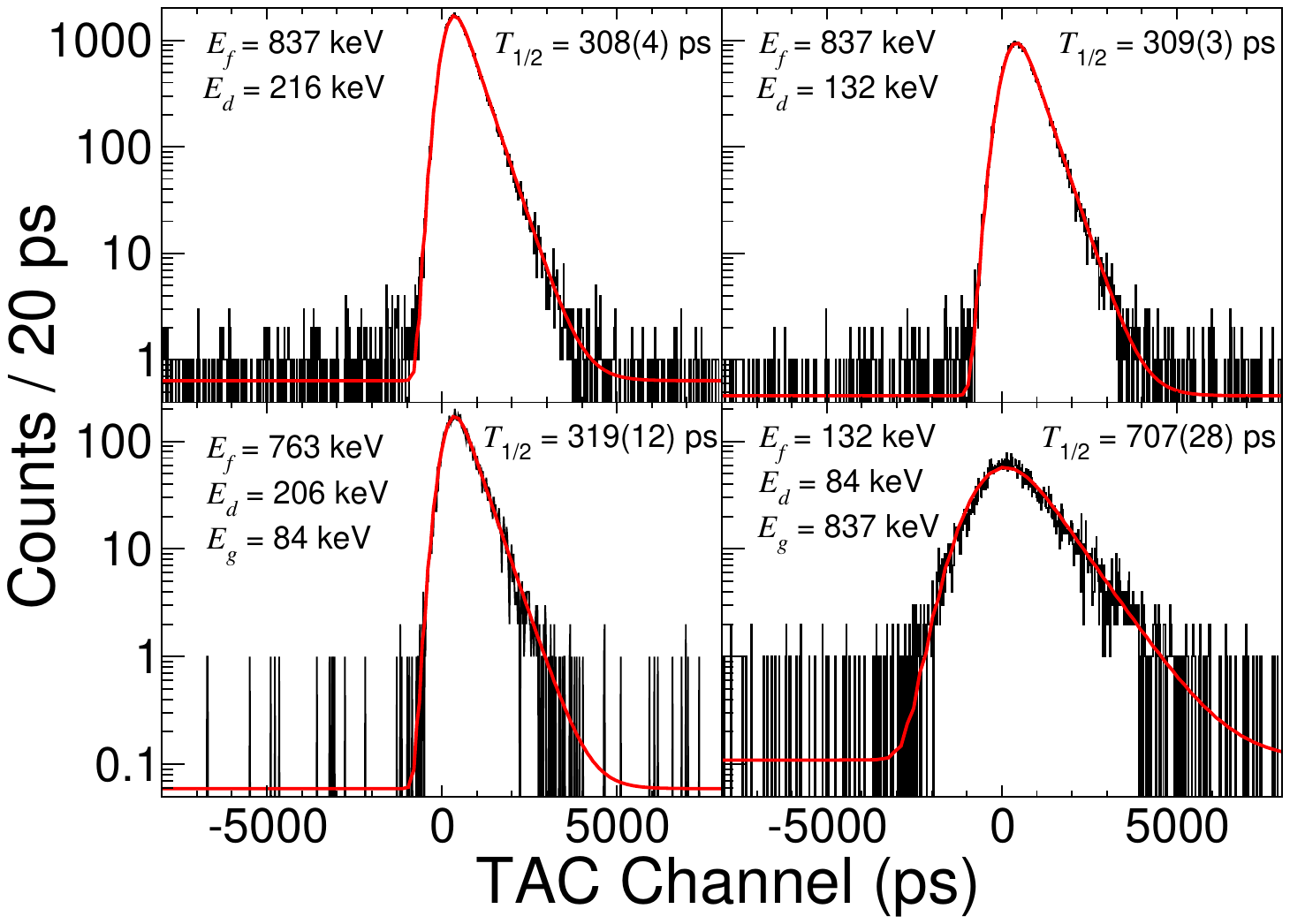}
    \caption{The combined time distributions of the delayed and anti-delayed histograms for various states measured in this work. Moving clockwise from the top left panel, the histograms represent the time distributions measured for the $J^\pi_i = 1^-_1$, $1^-_1$, $2^+_1$ and $3^-_1$ states. In each panel, the feeding, decaying and gating transitions used to generate the histogram are indicated with $E_f$, $E_d$ and $E_g$, respectively. The measured half-life, $T_{1/2}$, is also presented.}
    \label{TACspectra}
\end{figure*}

To confirm the methodology was sound, the lifetime of the first $J^\pi = 2^+$ state was measured and compared with values in the literature. This was performed by minimising background in the LaBr$_3$(Ce) spectra by demanding coincidences with the 837~keV transition, detected in the HPGe detectors, which connects the $1^-$ state at an excitation energy of 1053~keV and the $1^-$ state observed at 216~keV. The time differences observed between the 132~keV transition feeding the $2^+$ state and the 84~keV transitions decaying from this state are shown in Figure~\ref{TACspectra}. This spectrum was constructed by combining both the histograms corresponding to the delayed and anti-delayed configurations. The result of fitting the histogram in Figure~\ref{TACspectra} is the extraction of a mean lifetime corresponding to $\tau(2^+) = 1.02(4)$~ns. The uncertainty associated with this value is dominated by a systematic uncertainty which results from the presence of x-rays, characteristic to radium, which have a similar energy to the 84~keV $\gamma$ transition. Peaks corresponding to these x-ray transitions can be seen in the inset of Figure~\ref{labrSpec}. In demanding coincidences with the detection of 837~keV $\gamma$ rays in the HPGe detectors, the pathways to feeding the $2^+$ state were limited and therefore the influence of the x-rays on the measurement minimized. The present measurement of the lifetime of the $2^+$ state is lower than the value measured in previous studies~\cite{Neal_Kraner,Ton}, but remains statistically consistent with the adopted $1079(27)$~ps~\cite{A224} value. 

In the case of the measurement of the mean lifetime of the $1^-$ state, there was no need for an additional gate since the obtained spectra were sufficiently free of background due to the relatively high intensity of the feeding and decaying transitions. Figure~\ref{TACspectra} shows a histogram resulting from 837~keV ($E_f$) and 216~keV ($E_d$) gate combinations. Fitting the data of Figure~\ref{TACspectra} allows for the extraction of a mean lifetime corresponding to $\tau(1^-) = 444(6)$~ps. A second measurement of the lifetime of this state was also possible by using the 837~keV feeder and the 132~keV decay transitions. The histogram corresponding to this gate combination is shown in Figure~\ref{TACspectra}. The mean lifetime extracted from a fit of this histogram resulted in a value of $446(4)$~ps in excellent agreement with the other value. 

A similar analysis procedure was applied to extract a measurement of the mean lifetime of the $J^\pi = 3^-$ state. In this case, however, the intensity of the decaying transitions was significantly weaker and coincidences with the 84~keV transitions emitted in the decay of the first $2^+$ state were demanded. In doing so, the influence of the more intense 216~keV transition was minimised since this transition is not in coincidence with the 84~keV transition. However, this 84~keV $E2$ transition is expected to be highly converted ($\alpha = 21.6(3)$~\cite{briccPaper}) meaning that a significant fraction of its intensity was observed in the PACES detectors. Therefore, the TAC spectrum of Figure~\ref{TACspectra} represents the time difference between feeding ($E_f = 763$~keV) and decaying ($E_d = 206$~keV) transitions in coincidence with either 84~keV $\gamma$ rays detected with the HPGe detectors or conversion electrons (67 or 80~keV corresponding to $L$- and $M$-shell conversion electrons) detected in the PACES detectors. A spectrum showing the $L$- and $M$-shell conversion electrons in coincidence with the detection of 206~keV transitions in the LaBr$_3$(Ce) detectors is shown in the inset of Figure~\ref{pacesSpec}. Fitting the histogram in Figure~\ref{TACspectra} yielded a value of $\tau(3^-) = 460(18)$~ps. 

%\section{Results}

%\subsection{Lifetime measurements}

%\section{Discussion}
The lifetimes measured in the present work allow the reduced electric-dipole transition probabilities to be extracted for the key decays from the low-lying negative-parity states in $^{224}$Ra. Using the measured mean lifetimes, the observed branching ratios, and the appropriate internal-conversion coefficients~\cite{briccPaper}, the transition strengths were determined and are listed in Table~\ref{dipoleTable} alongside the values reported by Gaffney {\it et al}.~\cite{Gaffney}. The present values are in good agreement with those of Ref.~\cite{Gaffney}, but determine the relevant $E1$ strengths with significantly higher precision. This agreement is an important result since the two approaches are sensitive to the same underlying matrix elements but rely on very different experimental observables and analysis procedures and, hence, provide independent validation of the results of the Coulomb excitation experiment.

\begin{table*}
\centering
\caption{Table showing some properties of key states and transitions measured in this work including the $\gamma$-ray branching ratios ($b_\gamma$), reduced transition probabilities ($B(E\lambda)$) and intrinsic electric dipole and quadrupole moments ($Q_0^\lambda$). The relevant reduced transition probability values as measured by Gaffney {\it et al.}~\cite{Gaffney} are provided for comparison.}
\begin{tabular}{c c c c c c c c}
\hline
$J^{\pi}$ & $\tau (ps) $ & $J_i \rightarrow J_f$ & $E_d $ (keV) & $b_\gamma (\%)$ & $B(E\lambda)$ (W.u.) & $B(E\lambda)$ (W.u.)~\cite{Gaffney} & $Q_0^\lambda$ (efm$^\lambda) $\\
\hline
$2^{+}$ & $1020 \pm 40$ & $2^+ \rightarrow 0^+$ & 84 & $100$ & $105 \pm 4$ & $98 \pm 3$ & $652 \pm 26$ \\
$1^{-}$ & $444 \pm 6$ & $1^- \rightarrow 0^+$ & 216 & $68 \pm 3$ & $(3.5 \pm 0.1) \times 10^{-5}$ & $<5.0 \times 10^{-5}$ & $0.033 \pm 0.001$ \\
$1^{-}$ & $446\pm4$ & $1^- \rightarrow 2^+$ & 132 & $32 \pm 2$ & $(7.4 \pm 0.4) \times 10^{-5}$ & $<1.3 \times 10^{-4}$ & $0.033 \pm 0.001$ \\
$3^{-}$ & $460\pm18$ & $3^- \rightarrow 2^+$ & 206 & $98.2 \pm 0.7$ & $(3.6 \pm 0.2) \times 10^{-5}$ & $3.9^{+1.7}_{-1.4} \times 10^{-5}$ & $0.029 \pm 0.001$ \\
$3^{-}$ &  & $3^- \rightarrow 1^-$ & 74 & $1.8 \pm 0.1$ & $98 \pm 4$ & $93 \pm 9$ & $556 \pm 11$ \\
\hline
\end{tabular}
\label{dipoleTable}
\end{table*}

The improved precision is particularly valuable because the $E1$ strengths in this nucleus are extremely small in absolute terms, despite the strong octupole correlations implied by the low excitation energies of the $1^-_1$ and $3^-_1$ states and the previous measurement of a large $B(E3)$ strength~\cite{Gaffney}. The three measured $B(E1)$ values are mutually consistent with a common, strongly-quenched intrinsic dipole moment. In a simple $K = 0$ rotational description, the expected ratio $\frac{B(E1;1^-_1 \rightarrow 2^+_1)}{B(E1;1^-_1 \rightarrow 0^+_1)}$ is 2, which is consistent with the experimentally observed value of 2.11(13). In contrast, the measured ratio $\frac{B(E1;3^-_1 \rightarrow 2^+_1)}{B(E1;1^-_1 \rightarrow 0^+_1)} = 1.03(6)$ is significantly lower than the rigid-rotator expectation of $9/7 \simeq 1.29$. This discrepancy may indicate that the assumption of a spin-independent $E1$ matrix element is not satisfied between these members of this $K^\pi = 0^-$ rotational band. The corresponding intrinsic dipole moments extracted from the three transitions were calculated and can be seen in Table~\ref{dipoleTable}. Thus, despite the well-established octupole collectivity in $^{224}$Ra, the electric-dipole degree of freedom is strongly suppressed. This can be compared with the corresponding values reported for $^{228}$Th of 0.16(7) and 0.09(2)~$e$fm for the $1^-$ and $3^-$ states, respectively~\cite{Chishti2020}.

The present results may be compared with several theoretical approaches. Reasonable agreement is obtained between the experimental $B(E1)$ values measured in this work and the calculated values reported in Refs.~\cite{Robledo2011,Robledo2013}. In particular, the one-dimensional calculations, involving only the octupole degree of freedom, predicted values of $1 \times 10^{-5}$~W.u. and $7 \times 10^{-5}$~W.u using the D1S and D1M parameterisations, respectively. The two-dimensional calculations, which coupled the quadrupole and octupole degrees of freedom, significantly overestimate the $B(E1)$ strength with a value of $2.4 \times 10^{-4}$~W.u.~\cite{Robledo2013}.

Calculations based on the interacting boson approximation provide a complementary comparison. The mapped IBM calculations of Nomura~\cite{Nomura2025} predict substantially larger $E1$ strengths for the same low-lying transitions. For $^{224}$Ra, the calculated $sdf$-IBM values are $4.5 \times 10^{-3}$, $5.8 \times 10^{-3}$ and $6.4 \times 10^{-3}$~W.u. for the $1^-_1 \rightarrow 0^+_1$, $1^-_1 \rightarrow 2^+_1$ and $3^-_1 \rightarrow 2^+_1$ transitions, respectively. The corresponding $spdf$-IBM values are $2.4 \times 10^{-3}$, $4.0 \times 10^{-3}$ and $3.1 \times 10^{-3}$~W.u. The inclusion of the $p$-boson degree of freedom therefore reduces the predicted $E1$ strengths, but the calculated values remain larger than the present measurements by factors of approximately 50--150. This suggests that, while the mapped IBM approach captures important aspects of the quadrupole-octupole collective structure, it does not quantitatively reproduce the unusually strong quenching of the intrinsic dipole moment in $^{224}$Ra.

The small value of $D_0$ extracted in the present work is also consistent with the predictions of Butler and Nazarewicz~\cite{Butler1991}. In their macroscopic-microscopic description, the intrinsic dipole moment in reflection-asymmetric nuclei is decomposed into macroscopic and shell-correction contributions. For $^{224}$Ra, the macroscopic and shell contributions are approximately $+0.15$ and $-0.14~e{\rm fm}$, respectively, yielding a total intrinsic dipole moment close to zero, of order $D_0 \sim 0.01~e{\rm fm}$. The present value, $D_0 \simeq 0.032~e{\rm fm}$, is somewhat larger than this prediction, but remains very small on the scale of intrinsic dipole moments in octupole-correlated actinides. The agreement is therefore strongest at the qualitative level: both experiment and theory agree that the intrinsic dipole moment in $^{224}$Ra is strongly quenched by a near cancellation of larger opposing contributions.

The present data therefore reinforce the interpretation of $^{224}$Ra as a nucleus with strong octupole correlations, but an anomalously small intrinsic electric-dipole moment. The agreement with the Coulomb-excitation results of Gaffney {\it et al}.~\cite{Gaffney} validates the earlier matrix-element determination, while the improved precision of the lifetime measurements provides a much more stringent benchmark for theory. The comparison with microscopic EDF calculations~\cite{Robledo2013}, mapped IBM calculations~\cite{Nomura2025} and the macroscopic-microscopic predictions of Butler and Nazarewicz~\cite{Butler1991} shows that reproducing the low-lying negative-parity structure is not sufficient: the $E1$ transition strengths provide a particularly sensitive test of the delicate proton-neutron cancellation responsible for the observed dipole-moment quenching. This is particularly important for searches for non-zero electric dipole moments, since reliable calculations of Schiff moments in octupole-correlated nuclei require theoretical models that can reproduce not only excitation energies and collective octupole strength, but also the subtle proton-neutron cancellations revealed by the present $E1$ transition strengths.

%\section{Summary}

\begin{acknowledgments}

We would like to thank the beam delivery staff at TRIUMF for providing the radioactive beam. The GRIFFIN infrastructure has been funded jointly by the Canada Foundation for Innovation (CFI), the British Columbia Knowledge Development Fund (BCKDF), the Ontario Ministry of Research and Innovation (ON-MRI), TRIUMF and the University of Guelph. TRIUMF receives federal funding via a contribution agreement through the National Research Council Canada (NRC). This work was supported in part by the Natural Sciences and Engineering Research Council of Canada (NSERC). D.W., D.O., and M.B. would like to acknowledge funding from the UK Science and Technologies Facilities Council (STFC) with project codes ST/V001124/1 and ST/Y000382/1. 

\end{acknowledgments}

\bibliography{224RaLifetimes}

\end{document}